\documentclass[aps,showpacs,nofootinbib]{revtex4}
\usepackage{epsf,latexsym}
\usepackage{amsfonts}
\usepackage{graphicx}

\begin{document}

\title{Expectation values of minimum-length Ricci scalar}
\author{Alessandro Pesci\footnotetext{e-mail: pesci@bo.infn.it}}
\affiliation
{INFN Bologna, Via Irnerio 46, I-40126 Bologna, Italy}

\begin{abstract}
In this paper,
we consider a specific model,
implementing the existence of a fundamental limit 
distance $L_0$ between (space or time separated) points 
in spacetime,
which 
in the recent past 
has exhibited
the intriguing feature
of having
a minimum-length Ricci scalar $R_{(q)}$
that does not approach the ordinary Ricci scalar $R$
in the limit of vanishing $L_0$.
$R_{(q)}$ at a point
has been found to depend on the direction
along which the existence of minimum distance is
implemented. 

Here, we point out that
the convergence $R_{(q)}\to R$ in the $L_0\to 0$
limit is anyway recovered in a relaxed or generalized sense,
which is when we average over directions,
this suggesting we might be taking
the expectation value of $R_{(q)}$ promoted
to be a quantum variable.

It remains as intriguing as before
the fact that we cannot identify
(meaning this is much more than simply equating in the generalised sense above)
$R_{(q)}$ with $R$ in the $L_0\to 0$ limit,
namely when we get ordinary spacetime.
Thing is like if,
even when $L_0$ (read here the Planck length)
is far too small
to have any direct detection of it feasible,
the intrinsic quantum nature of spacetime
might anyway be experimentally at reach, 
witnessed by the mentioned special feature 
of Ricci, 
not fading away with $L_0$
(i.e. persisting when taking the $\hbar \to 0$ limit).

\end{abstract}


\maketitle

\vspace {1cm}
\noindent
In memory of Prof. Thanu Padmanabhan

$ $
\section{Introduction}

In the general task of finding a coherent description
of gravity and quantum mechanics,
methods playing 
on the existence of a minimum, or zero-point, length
(generically expected from 
the combination of gravitational and quantum 
basic requirements)
have proved to be of some value.  
Definite results have been obtained
through introduction of a quantum effective metric,
also called qmetric $q_{ab}$ \cite{KotE, Pad01, KotI},
characterized by the property
of providing a finite limit geodesic distance $L_0$ between
two points $P$ and $p$ in the coincidence limit $p\to P$
(thus $q_{ab}$ is clearly not a metric in any ordinary sense).
Examples of these are the extraction of entropy density
of spacetime \cite{Pad01, Pad02}, the indication that spacetime
is 2-dim at Planck scale \cite{Pad05} and a modified
Raychaudhuri equation \cite{KotG, ChaD}.

Of specific interest
for this paper
is the collection of results
regarding the qmetric Ricci scalar $R_{(q)}$ \cite{Pad01, KotI, PesP}.
A striking feature of them
is that the expression one gets for the latter
does not approach, in the limit $L_0 \to 0$, 
the Ricci scalar $R$ of ordinary metric.
Moreover,
the limit
$\lim_{p\to P} R_{(q)}$
with $R_{(q)}$ defined along a geodesic $\gamma$ from $P$, 
turns out to depend on $\gamma$, meaning on the direction
of approach to $P$.

The latter fact raises the very basic issue of
how can we have $R_{(q)}$ defined at a point $P$
if the value we ought to assign to it changes with the geodesic along
which $R_{(q)}$ is defined near $P$.
If we insist in regarding $R_{(q)}$, 
as probed by picking one of these geodesics at $P$,
as expressing {\it the} Ricci scalar at $P$ in the quantum metric
of that geodesic,
we need means by which the mentioned dependence on the geodesic
can result after the probe is only apparent or fictitious.
This has been tentatively explored in \cite{PesQ},
where it is discussed that there might be a sense in which
this can be accomplished,
if we endow matter with the capability to affect the geometry of spacetime
with the latter constrained by the matter content
through suitable field equations.
In other words,
in this perspective one would have to consider 
the microstructure of spacetime not {\it per se}
but in connection with dynamics. 

In this paper,
we try to face the issue with a complementary approach.
We would like to regard the qmetric Ricci scalar at $P$
as an elaborate object, defined through 
concurrent consideration of all geodesics through $P$.
The uncertainty principle itself suggests   
also such an attitude after all: 
When we try to probe curvature at $P$
at smaller and smaller scales, we get more and more uncertain
about momentum,
and we loose track of the actual curve along which 
we eventually reach $P$. 
Our goal here would be to leverage on the explicit expressions found for
$R_{(q)}$ based on single geodesics, 
to try to gain an understanding of the nature of this object
as a whole.

\section{Riemannian geometry}

As a preparation to consider the qmetric, or quantum, case,
we recall what the ordinary or classical metric says
about curvature when going very close to $P$.
We know actually, 
that in the minimum-length framework
this description is doomed to fail when we are close enough
to $P$.
However,
the behaviour when we are just about to be too close
for the classic description to be valid, 
might be hoped to provide indeed 
some precious hints on what this description might convert into.  

Two formulae we have at disposition
do relate curvature to areas of equigeodesic hypersurfaces
of congruences of geodesics emerging from $P$.
One is

\begin{eqnarray}\label{44.1}
\frac{A_P}{A_{P, flat}}(r) 
=
1 - \frac{R}{6 D} \, r^2 + {\cal O}(r^4)
\end{eqnarray}
\cite{GraA}
(assuming to be in a
Riemannian manifold $M$
with dimension $D$).
$A_P(r)$ is the ``area'' of the equigeodesic hypersurface 
$\Sigma(P, r)$
at geodesic distance $r$ from $P$. 
$A_{P, flat}(r)$ is the same were
the $D$-dimensional manifold flat Euclidean.

The other one is

\begin{eqnarray}\label{44.3}
\frac{dS_P(s, t^a)}{dS_{P, flat}(s)}
=
1 - \frac{1}{6} \, s^2 \, R_{ab} t^a t^b + {\cal O}(s^3)
\end{eqnarray}
\cite{VisA},
with $R_{ab}$ the Ricci tensor.
Here 
$M$ is a
($D$-dim) spacetime
and we are considering a congruence of
geodesics from $P$
with tangent $t^a = dx^a/ds$ with the parameter $s$ affine.
$t^a$ is a unit vector in case of a timelike or spacelike congruence
($s$ is geodesic distance defined as positive semi-definite).
In case of a timelike or a spacelike congruence,
$dS_P(s, t^a)$ denotes the ``area''
of the $(D-1)$-dim section at $s$ from $P$
of a narrow bundle of geodesics
emanating from $P$ around an assigned direction $t^a$ at $P$.
In case of null congruence,
$dS_P(s, t^a)$ is the ``area'' of $(D-2)$-dim spatial section
at affine parameter $s$ from $P$ of the bundle around $t^a$.
$dS_{P, flat}(r)$ is, in the two cases,
the same as (corresponding) $dS_P(s, t^a)$ were
the $D$-dimensional manifold flat Minkowski. 

Formula (\ref{44.3}) holds true of course
for Riemannian manifolds, 
in which case it can be written as

\begin{eqnarray}\label{44.3.bis}
\frac{dA_P(r, t^a)}{dA_{P, flat}(r)}
=
1 - \frac{1}{6} \, r^2 \, R_{ab} t^a t^b + {\cal O}(r^3),
\end{eqnarray}
with
$dA_P(r, t^a)$ the element of ``area'' of 
$\Sigma(P, r)$ at a point on $\Sigma(P, r)$
with corresponding geodesic tangent $t^a$
at $P$,
and
$dA_{P, flat}(r)$ is the same were
the $D$-dim manifold flat Euclidean.

Focusing on the Riemannian case,
we see that formula (\ref{44.3.bis}) is sort of
differential version (in solid angle) of
formula (\ref{44.1}).
Considering these formulae with $r$ very small,
this clearly implies that $R$ in (\ref{44.1})
has the meaning of kind of angular average
of the $R_{ab} t^a t^b$ terms of (\ref{44.3.bis}).  

Indeed,
let us fix a local Euclidean frame at $P$
with an orthonormal basis $\{{e^a}_i\}$ for it
(``$i$'' labels the vector, ``$a$'' the components)
in which $R_{ab}$ is diagonal
(that this can be done is guaranteed from being $R_{ab}$ symmetric). 

We have
$
t^a = \sum_{i=1}^{D} c_i \, {e^a}_i
$
(here and hereafter, 
no sum convention is applied on repeated indices
when they are tagging vectors 
(sum convention when they simply tag components)),
with the coefficients $c_i$ such that
$\sum_{i=1}^{D} {c_i}^2 = 1$.
And we get
$R_{ab} t^a t^b = 
R_{ab} \, \sum_{i=1}^{D} c_i \, {e^a}_i 
\, \sum_{j=1}^{D} c_j \, {e^b}_j =
\sum_{i=1}^{D} {c_i}^2 \, R_{ii}$.
Thus

\begin{eqnarray}\label{49.7}
\int R_{ab} t^a t^b \, d\Omega_{(D)}
&=&
\sum_{i=1}^{D} R_{ii}
\int {c_i}^2 \, d\Omega_{(D)}
\nonumber \\
&=&
\frac{1}{D} \, \sum_{i=1}^{D} R_{ii} \int d\Omega_{(D)}
\nonumber \\
&=&
\frac{1}{D} \, R \int d\Omega_{(D)}.
\end{eqnarray}
Here
$d\Omega_{(D)}$ denotes the element of solid angle
in $D$-dim Euclidean space,
and
the second equality in the above comes from 
$\int {c_i}^2 \, d\Omega_{(D)}
= 
\int (t_a \, {e^a}_i)^2 \, d\Omega_{(D)}
=
\int (t_a \, {e^a}_j)^2 \, d\Omega_{(D)}
=
\int {c_j}^2 \, d\Omega_{(D)}$,
$\forall i,j$, 
and from
$\int \sum_{i=1}^{D} {c_i}^2 \, d\Omega_{(D)}
=
 \int d\Omega_{(D)}$.

We have then

\begin{eqnarray}\label{46.3}
R 
=
\frac
{\int D \, R_{ab} t^a t^b \, d\Omega_{(D)}}
{\int d\Omega_{(D)}}.
\end{eqnarray}
This equation tells us 
that in a Riemannian manifold
the Ricci scalar does coincide with the mean 
over the direction of the quantity $D \, R_{ab} t^a t^b$.
In other words, the latter quantity can be interpreted
as ``density of Ricci scalar per unit solid angle in direction $t^a$ ''.

All this is seemingly
quite standard practice.
Point is now however
that the term
$D \, R_{ab} t^a t^b$ is precisely
the kind of quantity one gets in the
expression of the qmetric Ricci scalar $R_{(q)}$.

Indeed,
what has been found in \cite{KotI} is that
considering the qmetric $q_{ab}(p, P)$ associated in a spacetime
to a timelike or spacelike geodesic
($p$ runs along a timelike or spacelike geodesic from $P$;
$q_{ab}$ gives a new geodesic distance $\tilde s(p, P)$,
with ${\tilde s} \to L_0$ when $p \to P$) 
one has

\begin{eqnarray}\label{47.0}
R_{(q), t^a, L_0}
= 
\epsilon \, D \, R_{ab} t^a t^b + {\cal O}(L_0)
\end{eqnarray}
($\epsilon \equiv t^a t_a \pm 1$),
giving
$R_{(q), t^a, L_0} \to \epsilon \, D \, R_{ab} t^a t^b \, (\ne R)$
for $L_0\to 0$.
Here  
$
R_{(q), t^a, L_0} \equiv
\lim_{p\to P} R_{(q)}(p, P)
$
is the coincidence limit $p\to P$
of the qmetric Ricci 
(bi)scalar $R_{(q)}(p, P)$ (it depends on both $p$ and $P$),
namely of the Ricci scalar corresponding, 
at any $p$ on the geodesic, to $q_{ab}$.
In practice, 
$R_{(q), t^a, L_0}$ is the qmetric Ricci scalar
we may associate to point $P$ based on
geodesic of tangent $t^a$ at $P$.
In a Riemannian manifold, we get

\begin{eqnarray}\label{47.1}
R_{(q), t^a, L_0}
= 
D \, R_{ab} t^a t^b + {\cal O}(L_0),
\end{eqnarray}
with
$R_{(q), t^a, L_0} \to D \, R_{ab} t^a t^b$
for $L_0\to 0$.

When contrasted with (\ref{46.3}),
this result suggests that
the effect of the qmetric based on geodesic with  tangent $t^a$
may be regarded as to replace
the ordinary Ricci scalar $R$
with an effective Ricci scalar $R_{(q), t^a, L_0}$
(different from $R$ also in the $L_0\to 0$ limit)
with the latter having in the ordinary metric 
the interpretation (for $L_0$ small) of
``density of Ricci scalar per unit solid angle in
the direction $t^a \,$''.

Indeed,
writing
$R_{(q), t^a} \equiv \lim_{L_0\to 0} R_{(q), t^a, L_0}$,
we have

\begin{eqnarray}\label{47.5}
\frac{\int R_{(q), t^a} \, d\Omega}
{\int d\Omega} = R.
\end{eqnarray}
Since $d\Omega$ is the same in the ordinary metric and in the qmetric,
we see that the average over geodesic' direction of the ($L_0\to 0$) qmetric
Ricci scalar is the ordinary Ricci scalar.  

As a first result we have thus the statement that
for a Riemannian manifold,
even though the qmetric Ricci scalar based 
on geodesic with tangent $t^a$ goes in the $L_0\to 0$ limit
to something different
from the ordinary Ricci scalar $R$,
its average over $t^a$ does go to $R$.

All this can be re-expressed in a different, simple manner.
Let us take a local Euclidean frame at $P$
with orthonormal basis $\{{e^a}_i\}$ 
(in which $R_{ab}$ is generic,
nonnecessarily diagonal). 
We have

\begin{eqnarray}\label{54.1}
\sum_{i=1}^{D} R_{ab} \, {e^a}_i \, {e^b}_i
&=& 
\sum_{i=1}^{D} R_{ii}
\nonumber \\
&=&
R
\end{eqnarray}
(sum implied only on components' indices).
Then

\begin{eqnarray}\label{54.2}
\frac{1}{D} \, 
\sum_{i=1}^{D} D \, R_{ab} \, {e^a}_i \, {e^b}_i 
= R,
\end{eqnarray}
this giving the Ricci scalar $R$
as average of the quantity $D \, R_{ab} t^a t^b$
over the basis vectors ${e^a}_i$,
each ${e^a}_i$ coming with equal weight $1/D$.
Continuing with the terminology above,
we can think of $R_{ab} \, {e^a}_i \, {e^b}_i$
as the ``probability density of $R$ along
the $i$th orthogonal direction''.

From qmetric expression (\ref{47.1}), this is

\begin{eqnarray}\label{54.7}
\frac{1}{D} \,
\sum_{i=1}^{D} R_{(q), {e^a}_i}
= R.
\end{eqnarray}

The result we mentioned above can then be stated also as follows.
The average of the qmetric quantity
$R_{(q), t^a}$ at $P$ 
(i.e. of the $L_0\to 0$,  $p\to P$ limit
of the qmetric Ricci scalar based on
the geodesic with unit tangent $t^a$) 
over 
the basis vectors of the space swept
by the geodesics from $P$
is the ordinary Ricci scalar $R$.

We see, what we have is sort of expectation value,
one might think, 
of a quantum variable $R_{(q)}$, the qmetric Ricci scalar at $P$,
in the space of possible directions.
As hinted above, 
the picture would then be supposed to be that, 
as soon as we explore
space 
closer and closer to $P$,
we become completely uncertain about the geodesics 
on which we are,
and the quantum variable $R_{(q)}$
has equal probability
to be probed along any direction.
This assigns to each measurement result $R_{(q), {e^a}_i}$,
i.e. to what we obtain if we happen to be along 
any of basis vectors ${e^a}_i$, 
equal probability $1/D$.

\section{Lorentzian geometry}

We proceed now to take $M$ a ($D$-dim) spacetime.
In the Lorentzian case,
in general we do not have at disposition something
like (\ref{44.1}).
As a consequence, it is not clear in principle
which meaning we can give, if any, 
to taking kind of ``average on direction'',
somehow along what we actually did in the Riemannian case.

At the end of previous Section however, 
we saw that the Ricci scalar $R$ can conveniently be
expressed as average 
over the basis vectors of the quantity (for Riemannian case) 
$D \, R_{ab} t^a t^b$ (equation (\ref{54.2})).
This can straightforwardly be exported to Lorentzian case.

Indeed,
let us set up a local Lorentz frame at $P$
with orthonormal basis $\{{e^a}_i\}$, 
where $i = 0, 1, ..., D-1$ labels the vector
and ${e^a}_0$ is timelike.
In it,
$t^a = \sum_{i=0}^{D-1} c_i \, {e^a}_i$
with
$- {c_0}^2 + \sum_{i=1}^{D-1} {c_i}^2 = -1, \, 0, \, 1$
(using for the metric the signature $(-, +, +, ...)$).

Considering the quantity
$Q(t^a) \equiv \epsilon(t^a) \, R_{ab} t^a t^b$,
where
$\epsilon(t^a) \equiv t^a\, t_a \, (= \pm 1, \, 0)$,
we have

\begin{eqnarray}\label{54.3}
\sum_{i=0}^{D-1} Q({e^a}_i)
&=&
\sum_{i=0}^{D-1} \epsilon_i \, R_{ab} \, {e^a}_i \, {e^b}_i
\nonumber \\ 
&=&
- R_{00} + \sum_{i=1}^{D-1} R_{ii} 
\nonumber \\
&=&
R,
\end{eqnarray}
with
$\epsilon_i \equiv \epsilon({e^a}_i)$. 

This equation can be recast as

\begin{eqnarray}\label{54.4}
\frac{1}{D} \, 
\sum_{i=0}^{D-1} \epsilon_i D \, R_{ab} \, {e^a}_i \, {e^b}_i
= R.
\end{eqnarray}
This shows that
in the Lorentzian case,
the ordinary Ricci scalar $R$
can be expressed as average 
of the quantity $\epsilon \, R_{ab} t^a t^b$
over the basis vectors.
Relation (\ref{54.4}) is the generalisation of (\ref{54.2}) 
to semi-Riemannian manifolds with index $\nu > 0$.




Clearly, we act this way and get this same result whichever is
the Lorentz frame we choose at $P$.
Indeed, we see this amounts in using the prescription
$t_a t_b \to \epsilon(t^c) \, g_{ab}$, 
averaging on $(D-1$)-dim Euclidean
hypersphere,
and continuing back the result to the Lorentzian metric;
it is the ``elementary averaging procedure'' of \cite{ALN}.
It corresponds to go from 
$t^a = (t^0, t^1, ..., t^{D-1})$
in the local Lorentz frame 
to 
$t^a_E = (t^0_E, t^1, ..., t^{D-1})
= (i t, t^1, ..., t^{D-1})$,
which, with $j, k = 1, ..., D-1$,
brings
\begin{eqnarray}
t^a \, R_{ab} \, t^b = 
(t^0)^2 \, R_{00} + 2 \, t^0  \, R_{0j} \, t^j 
+ t^j  t^k \, R_{jk}
\nonumber
\end{eqnarray}
to 
\begin{eqnarray}
t^a_E R_{ab} t^b_E &=& 
(t^0_E)^2 \, R_{00} + 2 \, t^0_E \, R_{0j} \, t^j 
+ t^j  t^k \, R_{jk}
\nonumber \\
&=& 
- (t^0)^2 \, R_{00} + 2 i \, t^0 \, R_{0j} \, t^j 
+ t^j  t^k \, R_{jk}.
\nonumber
\end{eqnarray}
We average now over angle on the (unit) $(D-1)$-sphere,
and we get
\begin{eqnarray}
\langle t^a_E \, R_{ab} \, t^b_E\rangle  
&=&
\langle (t^0_E)^2 \rangle \, R_{00} 
+ 2 \, \langle t^0_E t^j \rangle \, R_{0j} 
+ \langle t^j \, t^k \rangle \, R_{jk}
\nonumber \\
&=&
\langle - (t^0)^2 \rangle \, R_{00} 
+ 2 i \, \langle t^0 t^j \rangle \, R_{0j} 
+ \langle t^j \, t^k \rangle \, R_{jk},
\nonumber
\end{eqnarray}
with
$\langle (t^0)^2 \rangle = 1/D = \langle (t^j)^2 \rangle$,
$\langle t^0 t^j \rangle = 0 = \langle t^j \, t^k \rangle$, 
$k \ne j$.
This is the average we use in (\ref{54.4}).

Since Lorentz transformations are rotations in the Euclidean sector
with the rotation angle matching the tilt angle of the axes in
the Lorentz transformation,
this is taking an angular average over all paths to $P$ 
in the local coordinates (cf. \cite{ALN}),
and the procedure assures 
that if we let the basis vectors ${e^a}_i$ undergo
an active Lorentz transformation,
the result of the sum 
in the right-hand side 
of the first equality of (\ref{54.3})
stays unchanged 
as it ought to be for a sensible averaging procedure.

This is entirely different 
from trying to average 
on equigeodesic hypersurfaces from $P$ 
(thing which we did in the Riemannian case) 
in Lorentz;
the reason being, as mentioned, 
the fact that in the Lorentzian case we do not have
something like (\ref{44.1}).
As can be easily verified, 
had we tried this kind of average we would have found divergent quantities
(the domain of integration is no longer finite), 
no possibility to reliably approximate the integrand as evaluated on the
hypersurface with its value at $P$ (presence of regions
at fixed small 
geodesic distance 
being at arbitrarily large spatial (or time) distance
in the local frame at $P$),
and, morover,  
the integration of the time component would have been 
in any case different
from that of the spatial components 
(and with a ``wrong'' sign) contrary to what happens
in the Riemannian case and exploited in (\ref{49.7}).

From the qmetric result (\ref{47.0}),
we see then that with (\ref{54.4}) 
we obtain again equation (\ref{54.7})
(in terms of $i \in \{0, 1, ..., D-1\}$)

\begin{eqnarray}\label{54.5}
\frac{1}{D} \,
\sum_{i=0}^{D-1} R_{(q), {e^a}_i}
= R,
\end{eqnarray}
since now
$R_{(q), t^a} \equiv \lim_{L_0\to 0} R_{(q), t^a, L_0} =
\epsilon \, D \, R_{ab} t^a t^b$.
The comments right below equation (\ref{54.7}) up to the end of Section II
do apply unchanged thus also for $D$-dim spacetime,
among them the fact that,
even though the quantity $R_{(q), t^a, L_0}$
(i.e. the limit $p\to P$ of the qmetric Ricci scalar
based on geodesics of tangent $t^a$) 
goes in the $L_0\to 0$ limit to something $\ne R$,
the average over $t^a$ does go to $R$.





\section{Null expectation values}

Having gained in previous section some understanding 
of the qmetric spacetime quantities 
$R_{(q), t^a} = \epsilon \, D \, R_{ab} t^a t^b$
as possible outcomes of the quantum variable $R_{(q)}$,
the qmetric Ricci scalar,
when probed approaching $P$,
our aim in this section is to find what 
this description becomes
if our approach to $P$ is constrained 
to happen through null geodesics.
The reason behind this further analysis
is the result \cite{PesP}
concerning the expression of the qmetric Ricci scalar
based on null geodesics from $P$.

Specifically,
from the expression of the qmetric $q_{ab}$ 
corresponding to the case of null separated events \cite{PesN},
what has been found in \cite{PesP} is that 

\begin{eqnarray}\label{riscaq}
R_{(q), \, l^a, L_0} = (D-1) \, R_{ab} l^a l^b + {\cal O}(L_0). 
\end{eqnarray}
Here $l^a = dx^a/d\lambda$ is the tangent at $P$ to a null geodesic $\gamma$
with affine parameter $\lambda$.
The considered qmetric $q_{ab}$ is that corresponding 
to the assigned null geodesic $\gamma$,
and
$R_{(q), \, l^a, L_0}$ is the coincidence limit $p\to P$ along $\gamma$,
of the qmetric Ricci biscalar $R_{(q)}(p, P)$ corresponding 
to $q_{ab}$.
In \cite{PesP},
the expression (\ref{riscaq}) of qmetric Ricci scalar has been
derived through consideration
of Gauss-Codazzi equations generalised \cite{Gem, ChaC_bis} to the
case of null hypersurfaces.
%
This expression applies when
the geometry of spacetime at $P$ happens to be such that
the consideration of a specific congruence of null geodesics 
in all directions from $P$
is enough to fix the Ricci scalar $R$ at $P$
(in spite of being the submanifold swept by these geodesics
only $(D-1)$-dim).

To see what the approach described above
gives in this case,
we come back to our local Lorentz frame at $P$
(with $R_{ab}$ nonnecessarily diagonal)
attached to a local observer at $P$,
and consider as $\gamma$'s the null geodesics from $P$
with affine parameter $\lambda$ given by distance according to that observer. 
Our aim is to gain some understanding of expression (\ref{riscaq})
for the qmetric associated to these $\gamma$'s
in terms of averages as above,
or, the other way around, to see if the criteria applied 
in previous sections  
allow us to predict the expression (\ref{riscaq}) for the null case.

Our first task is to try
to see if we can find an expression for $R$
like (\ref{54.1}) or (\ref{54.3}),
but this time in terms of null vectors.

We consider the set of $D-1$ null geodesics $\gamma_i$ from $P$,
parametrized by distance,
with tangent vectors ${l^a}_i$ at $P$ 
whose space components coincide in turn with
each spacelike vector of an orthonormal basis $\{{e^a}_i\}$,
$i = 0, 1, ..., D-1$,
where ${e^a}_0$ is the velocity of the observer.
At $P$ we introduce
the auxiliary null vectors 
${m^a}_i \equiv 2 \, {e^a}_0 - {l^a}_i$.
The couple of null vectors ${l^a}_i$ and  ${m^a}_i$ 
for a given $i$, span
the vector space generated by ${e^a}_0$ and ${e^a}_i$,
2-dim subspace of the tangent space $T_P(M)$.
 
Using the relations
${e^a}_0 = \frac{1}{2} \, ({l^a}_i + {m^a}_i)$
and
${e^a}_i = \frac{1}{2} \, ({l^a}_i - {m^a}_i)$,
$\forall i = 1, ..., D-1$,
we can express $R$ at $P$ in terms of the vectors
$\{{l^a}_i\}_{i=1, ..., D-1}$ and $\{{m^a}_i\}_{i=1, ..., D-1}$.
Starting from (\ref{54.3})
and calling $i'$ any of the numbers $\{1, ..., D-1\}$,
at $P$ we get

\begin{eqnarray}\label{55.1}
R 
&=&
\sum_{i=0}^{D-1} \epsilon_i \, R_{ab} \, {e^a}_i \, {e^b}_i
\nonumber \\
&=&
\frac{1}{4} \bigg[\epsilon_0 \, 
R_{ab} ({l^a}_{i'} + {m^a}_{i'}) ({l^b}_{i'} + {m^b}_{i'})
+ \sum_{i=1}^{D-1} \epsilon_i \, 
R_{ab} ({l^a}_i - {m^a}_i) ({l^b}_i - {m^b}_i)\bigg]
\nonumber \\
&=&
\frac{1}{4} \bigg[-\frac{1}{D-1} \, 
\sum_{i = 1}^{D-1} R_{ab} ({l^a}_{i} + {m^a}_{i}) ({l^b}_{i} + {m^b}_{i})
+ \sum_{i=1}^{D-1} 
R_{ab} ({l^a}_i - {m^a}_i) ({l^b}_i - {m^b}_i)\bigg]
\nonumber \\
&=&
\frac{1}{4} \bigg[ 
\sum_{i = 1}^{D-1} \frac{D-2}{D-1} \,
R_{ab} \, {l^a}_{i} \, {l^b}_{i}
+ \sum_{i = 1}^{D-1} \frac{D-2}{D-1} \,
R_{ab} \, {m^a}_{i} \, {m^b}_{i}
- 2 \, \sum_{i=1}^{D-1} \frac{D}{D-1} \,
R_{ab} \, {l^a}_i \, {m^b}_i \bigg]
\nonumber \\
&=&
\frac{1}{4} \,
\sum_{i=1}^{D-1} R_{ab} 
\bigg[
\frac{D-2}{D-1} \, \big({l^a}_i \, {l^b}_i + {m^a}_i \, {m^b}_i\big)
- 2 \, \frac{D}{D-1} \, {l^a}_i \, {m^b}_i \bigg],
\end{eqnarray}
where in the first sum of third equality
all the terms are equal.

This is equation (\ref{54.3}) when expressed in terms of null
vectors, and is as far we can go for spacetime generic at $P$.
We have now to express the special geometric circumstance 
that the geodesics sent out by our observer 
are actually able to fix the Ricci scalar $R$ at $P$.

Defining $L$ the submanifold swept 
($(D-1)$-dim, null) by all null geodesics of our congruence,
what we need is that $R$ be given with $L$,
namely that $L$ taken alone has in it all the information needed
to reconstruct the scalar curvature of $M$ at $P$
which, as mentioned above, are the circumstances to which
equation (\ref{riscaq}) refers.

From (\ref{44.3}), 
we see that terms of the sort
$R_{ab} \, {l^a}_i \, {l^b}_i$ have a meaning assigned with the congruence
(they express ratios of areas of bundles of geodesics;
terms like these are given with $L$,
even though $R_{ab}$ does depend on the characteristics of $M$ 
in a neighbourhood of any $p\in L$, thus also on what happens off $L$).
The vectors ${m^a}_i$ in (\ref{55.1}) at $P$,
do are tangent to geodesics of the congruence;
the same comment just expressed applies then to them as well.

We must find the conditions under which
the remaining terms in (\ref{55.1}),
namely the terms $R_{ab} \, {l^a}_i \, {m^b}_i$,
can be reduced too 
to terms of the kind $R_{ab} \, {l^a}_i \, {l^b}_i$
(or $R_{ab} \, {m^a}_i \, {m^b}_i$).

In our basis we have

\begin{eqnarray}\label{56}
R_{ab} \, {l^a}_i \, {l^b}_i
&=& 
R_{00} + 2 \, R_{i0} + R_{ii},
\nonumber \\
R_{ab} \, {m^a}_i \, {m^b}_i
&=& 
R_{00} - 2 \, R_{i0} + R_{ii},
\nonumber \\
R_{ab} \, {l^a}_i \, {m^b}_i
&=&
R_{00} - R_{ii}
\nonumber
\end{eqnarray}
at $P$.

Thus we see that the terms $R_{ab} \, {l^a}_i \, {m^b}_i$
reduce to (something proportional to) the other terms
when $R_{i0} = 0$ and, further,
$R_{00} = 0$ or $R_{ii} = 0$, $\forall i\in \{1, ..., D-1\}$. 

Of these two possibilities,
the latter should be disregarded since
any spatial rotation of our frame
would generate nonvanishing diagonal terms,
yet sending our congruence in itself.

We are left thus with 

\begin{eqnarray}\label{62.1}
R_{00} = 0, \, \, R_{i0} = 0, \, \, R_{ij} 
\, \, {\rm generic}, 
\, \, \forall i, j \in \{1, ..., D-1\}.
\end{eqnarray} 
This means that if the spacetime $M$ we are considering
does admit the existence at $P$ of an observer
for which this condition holds, 
the hypersurface $L$ swept  
by the congruence of geodesics sent out by this observer
has in it all what is needed to provide $R$ at $P$. 

Using this in the expression of $R$,
we get

\begin{eqnarray}\label{57.5}
R
&=&
\sum_{i=0}^{D-1} \epsilon_i \, R_{ab} \, {e^a}_i \, {e^b}_i
\nonumber \\
&=&
\sum_{i=1}^{D-1} R_{ab} \, {e^a}_i \, {e^b}_i
\nonumber \\
&=&
\frac{1}{4} \,
\sum_{i=1}^{D-1} 
\big(R_{ab} \, {l^a}_i \, {l^b}_i
+ R_{ab} \, {m^a}_i \, {m^b}_i
- 2 \, R_{ab} \, {l^a}_i \, {m^b}_i\big)
\nonumber \\
&=&
\sum_{i=1}^{D-1} R_{ab} \, {l^a}_i \, {l^b}_i. 
\end{eqnarray}

We have here the sum over the null vectors ${l^a}_i$
in all possible orthogonal spatial directions.
Being it a sum over $D-1$ possibilities,
we write

\begin{eqnarray}\label{57.7}
\frac{1}{D-1} \,
\sum_{i=1}^{D-1} (D-1) \, R_{ab} \, {l^a}_i \, {l^b}_i
= R.
\end{eqnarray}
This is the analogous for null rays 
of equations (\ref{54.2}) and (\ref{54.4}).
Equation (\ref{54.5}) becomes for them

\begin{eqnarray}\label{57.9}
\frac{1}{D-1} \, \sum_{i=1}^{D-1} R_{(q), \, {l^a}_i} = R,
\end{eqnarray}
with
$
R_{(q), \, {l^a}_i} \equiv \lim_{L_0 \to 0} R_{(q), \, {l^a}_i, \, L_0}$.

Thus,
for the mentioned geometric circumstances,
the average of the qmetric quantity $R_{(q), \, l^a}$ 
at $P$ (based on geodesic with affine parameter $\lambda$
and $dx^a/d\lambda = l^a$) 
evaluated over the null vectors ${l^a}_i$
corresponding to geodesics 
parametrized with distance
in all possible mutually orthogonal spatial directions 
gives $R$.  

The other way around,
if we know in advance
(for example on the basis
of timelike/spacelike results),
that the qmetric quantity $R_{(q), \, l^a}$
must be given the interpretation of ``probability density
of ordinary $R$ along direction $l^a \,$''
(this direction being meant as one of all the possible 
mutually spatially orthogonal 
null directions
in a frame in which $\lambda$, 
defined by $dx^a/d\lambda = l^a$,
is distance),
we can anticipate at once its expression to be
$R_{(q), \, l^a} = (D-1) \, R_{ab} \, l^a l^b$.

Further on,
were it known in advance for some reason that
the qmetric quantity $R_{(q), \, t^a}$,
with $t^a$ unit timelike or spacelike,
ought to be given the interpretation of
``probability density of $R$ along
direction $t^a \,$''
(this direction being meant as one of all possible 
mutually orthogonal directions),
from (\ref{54.4})
one could have anticipated at once the expression
$R_{(q), \, t^a} = \epsilon \, D \, R_{ab} t^a t^b$.

In \cite{PesP},
the geometric circumstance of $R$ 
being retrievable at $P$ from 
the specific null hypersurface $L$ taken alone, 
has been expressed as

\begin{eqnarray}\label{64.1}
V^a \nabla_a(v^b \nabla_b w^c) = 0
\end{eqnarray}
at $p\in L$, for all $v^a$, $w^a$ vector fields on $M$
given with the metric. $V^a$ is our ${e^a}_0$ here, parallel translated
along any $\gamma$ of the congruence.
Equation (\ref{64.1}) entails

\begin{eqnarray}\label{64.2}
{R^a}_{bcd} w^b v^c V^d = 0
\end{eqnarray}
at $p\in L$, $\forall v^a, w^a$ on $M$ given with the metric.

Applying this equation in our local Lorentz frame
to any vector of the basis,
at $P$ we get 
$
{R^a}_{bcd} \, {e^b}_{b'} \, {e^c}_{c'} \, {e^d}_0 = 0$,
with $b', c' \in \{0, 1, ..., D-1\}$ identifying the basis vector.
This is
${R^a}_{bc0} = 0$,
and
gives 
$R_{00} = 0$, $R_{i0} = 0$, 
$R_{ij} = {R^a}_{iaj}$ generic,
$i, j \in \{1, ..., D-1\}$,
in this frame at $P$;
conditions which coincide with (\ref{62.1}).

\section{Conclusions}

The striking feature of the minimum-length Ricci scalar
of not tending to the ordinary Ricci scalar $R$ 
when the limit length $L_0 \to 0$ is taken \cite{Pad01, KotI}, 
appears something worth
of further scrutiny.
What we have seen here,
is that the qmetric quantities
$R_{(q), t^a}$ ($\ne R$) at a point $P$ of a Riemannian space
or of spacetime,
namely, 
the limit when $L_0 \to 0$ of the Ricci scalar based on
qmetric from geodesic with tangent vector $t^a$at $P$, 
can be given in the starting ordinary space or spacetime 
the interpretation of sort of
``probability density of $R$ along $t^a \,$''.
What we mean with this
is that,
when we take the (unweighted) average of these quantities
over the set of allowed orthogonal directions,
we get $R$
(equations (\ref{54.2}-\ref{54.7}), 
(\ref{54.4}-\ref{54.5}) and (\ref{57.7}-\ref{57.9})).

This fits with an interpretation of the qmetric Ricci scalar
$R_{(q)}$ as kind of quantum variable.
$R$ can e.g. be considered
the expectation value 
$\langle R_{(q)} \rangle_\rho = {\rm Tr} \, \rho R_{(q)}$
of the quantum variable $R_{(q)}$
in the quantum maximally mixed state given by
$\rho = k \sum_i |{t^a}_i\rangle \langle {t^a}_i|$
($k >0$, normalisation constant),
where the $|{t^a}_i\rangle$'s 
form
a basis of mutually orthogonal
quantum states, corresponding to the allowed mutually orthogonal
directions ($a$ labels components, and $i$ the vector). 

Indeed,
from the uncertainty principle, 
we might expect that
the closer we get to $P$ the more uncertain we happen to be
concerning the geodesic from $P$ on which we are.
This assigns to every geodesic through $P$
an equal probability to be picked out
and thus an equal probability for projector
$|{t^a}_i\rangle \langle {t^a}_i|$ on each state $|{t^a}_i\rangle$
of the basis.

This shows that there is at least one sense in which
$R$ is recovered from its qmetric counterpart $R_{(q)}$
in the $L_0\to 0$ limit, 
namely,
in terms of average on orthogonal directions
or of expectation value of $R_{(q)}$ considered as quantum variable.

It remains as striking as before however,
the fact that this coincidence with $R$
in the $L_0\to 0$ limit, 
i.e. when we go to a situation in which space or spacetime
is no longer endowed with a limit length
(exactly what one gets in ordinary space or spacetime),
is not an identity, but is defined
in terms of averages or expectation values.
That is to say,
even when the limiting length becomes vanishingly small,
so that for example any possibility to measure it directly fades away,
what happens is that still we can get the Ricci scalar at $P$
only through averages or expectation values,
this possibly being sort of nonfading-away record  
of inherent quantum nature of space or spacetime
(cf. \cite{Pad01}),
persisting in the $\hbar\to 0$ limit.
This might provide (or add) some opportunity 
of detection of quantum gravitational features 
in spite of the smallness of Planck length scale.

The results here presented about the curvature of a spacetime endowed
with a minimum length might be connected with the notion of curvature
found in discrete approaches to quantum gravity.
Specifically,
equation (\ref{54.5}), arising from 
the qmetric description of the Ricci scalar in a generic spacetime
and a  probabilistic
interpretation, might be compared
with the notion of curvature 
as derived in causal set theory.
In the latter in fact,
the Ricci scalar of a 4-dim spacetime $M_{(4)}$
can be recovered
from quantities in the causal set generated by $M_{(4)}$,
and this offers a way to relate quantities 
in the two approaches.

In causal set theory
the Ricci scalar of the approximating manifold $M_{(4)}$ 
can be obtained
as the continuum limit
\begin{eqnarray}
\nonumber
R(P) =
\lim_{l\to 0} \,
4 \sqrt\frac{2}{3}
\, \frac{1}{l^2}
\, \langle \sum_{P'\preceq P} \delta_{P P'}
- \sum_{P'\in L_1(P)}
+ 9 \sum_{P'\in L_2(P)}
- 16 \sum_{P'\in L_3(P)}
+ 8 \sum_{P'\in L_4(P)} \rangle,
\end{eqnarray}
where $P, P'$ belongs to the causal set $\cal{C}$,
$\preceq$ is the partial order relation in $\cal{C}$,
$L_1(P)$ is the layer of the nearest neighbors of $P$ in its past, 
$L_2(P)$ the layer of the next to nearest neighbors,
and so on for $L_3(P)$ and $L_4(P)$
(\cite{Sorkin, BenincasaDowker, Benincasa},
see also \cite{Surya}).
$l$ is a length, and the average
is taken over many causal sets $\cal{C}$
each obtained
by stochastically selecting (sprinkling) points
from the manifold $M_{(4)}$ according to a same 
spacetime density $\rho = 1/l^4$.
If we then allow for the qmetric limit distance $L_0$ 
and the causal set discreteness scale $l$ to be
small together (each vanishing with the other),
from (\ref{54.5}) we get

\begin{eqnarray}\label{68.8}
&\frac{1}{4}& \sum_{i = 0}^3 R_{(q), {e^a}_i} =
\nonumber \\
\lim_{l\to 0}
&4\sqrt\frac{2}{3}&
\frac{1}{l^2} \,
\langle \sum_{P'\preceq P} \delta_{P P'}
- \sum_{P'\in L_1(P)}
+ 9 \sum_{P'\in L_2(P)}
- 16 \sum_{P'\in L_3(P)}
+ 8 \sum_{P'\in L_4(P)} \rangle,
\nonumber
\end{eqnarray}
which for $l$ small gives

\begin{eqnarray}\label{68.9}
\langle \sum_{P'\preceq P} \delta_{P P'}
- \sum_{P'\in L_1(P)}
+ 9 \sum_{P'\in L_2(P)}
- 16 \sum_{P'\in L_3(P)}
&+& 8 \sum_{P'\in L_4(P)} \rangle
\nonumber \\
&\sim&
\sum_{i = 0}^3 \epsilon_i \, l^2 \, R_{ab} \, {e^a}_i \, {e^b}_i
\nonumber \\ 
&=&
l^2 \, \langle R_{(q)} \rangle,
\end{eqnarray}
with the two averages having different meaning,
being the first taken over instances of causal sets
generated with a same density $\rho = 1/l^4$ 
from the manifold $M_{(4)}$,
and the second over probes of the qmetric Ricci scalar
along any direction at $P\in M_{(4)}$.

One last remark.
If we are given grounds for using
the interpretation of $R_{(q), t^a}$ as probability density
(in the meaning above) beforehand,
we can get the expression for $R_{(q), t^a}$
straight from equations (\ref{54.2}), (\ref{54.4}) or (\ref{57.7}).
In particular,
if we use the result (\ref{47.0})
for timelike and spacelike geodesics,
and equation (\ref{54.4}) to guess
the interpretation of $R_{(q), t^a}$ as probability density,
we get from (\ref{57.7}) an independent 
(from the derivation in \cite{PesP})
and 
quite economy way to derive
the expression (\ref{riscaq}).

{\it Acknowledgements.}
I wish to thank Sumanta Chakraborty and Dawood Kothawala
for consideration of an early version of the paper,
and the anonymous reviewer for a very helpful
observation concerning the discussion of Lorentz case.


\end{document}